\documentclass[pra,twocolumn]{revtex4}
\usepackage{graphicx}
\usepackage{dcolumn}
\usepackage{bm}

\begin{document}

\title{Personal Email Networks: An Effective Anti-Spam Tool}
\author{P. Oscar Boykin }
\author{Vwani Roychowdhury}
\email{boykin,vwani@ee.ucla.edu.}
\affiliation{Department of Electrical Engineering, University of California, Los Angeles,
CA 90095}
\date{February 3, 2004}

\begin{abstract}\small 

We provide an automated graph theoretic method for identifying individual
users' trusted networks of friends in cyberspace. We routinely use our
social networks to judge the trustworthiness of outsiders, i.e., to decide
where to buy our next car, or to find a good mechanic for it.  In this
work, we show that an email user may similarly use his email network,
constructed solely from sender and recipient information available in
the email headers, to distinguish
between unsolicited commercial emails, commonly called ``spam", and emails
associated with his circles of friends.
We exploit the properties of
social networks to construct an automated anti-spam tool which processes
an individual user's personal email network to simultaneously identify the
user's core trusted networks of friends, as well as subnetworks generated
by spams.  In our empirical studies of individual mail boxes, our
algorithm classified approximately 53\% {\em of all emails} as spam or
non-spam, with 100\% accuracy.
Some of the emails are left unclassified by this network analysis tool.
However, one can exploit two of the following useful features.
First, it requires no user intervention or supervised training;
second, it results in no false negatives i.e., spam being misclassified
as non-spam, or vice versa.
We demonstrate that these two features suggest that our
algorithm may be used as a platform for a comprehensive solution to the
spam problem when used in concert with more sophisticated, but more
cumbersome, content-based filters.
\end{abstract}

\maketitle

\section{Introduction \label{sec:intro}}
The amount of unsolicited commercial email (spam), and
more importantly, the fraction of email which is spam, has risen dramatically
in the last few years. Recently, a study has shown that
$52\%$ of email users say spam has made them less trusting
of email, and $25\%$ say that the volume of spam has
reduced their usage of email\cite{pipspam}.
This crisis has prompted 
proposals for a broad spectrum of potential solutions, ranging from
the design of more efficient anti-spam software tools to calls for 
anti-spam laws at both the federal and state levels. While the jury is 
still out on how widely such anti-spam laws will be 
enacted and how effective they would be in stemming the flow, the objective of 
the various legal and technical solutions are the same: to make it unprofitable 
to send spam and thereby destroy the spammers' underlying business model. 

Any measure that  stops spam from reaching users' inboxes with probability
$p$ and is deployed by users with probability $q$ increases
the average cost of sending spam by $1/(1-pq)$.
For instance, if a filter has 90\% accuracy and is used by 90\% 
of users, the the cost of sending spam will
increase by more than $500\%$.
However, even if a filter is 99.9\% accurate, but only 1\% of users
actually use it, then spammers' costs will only increase
by $1\%$.  Thus, the need is for anti-spam techniques that work accurately enough and, 
more importantly, are user friendly and computationally efficient
so that they can be widely deployed and 
used.  In evaluating the accuracy of the anti-spam 
tool, it is especially important that the algorithm should generate virtually 
no false negatives, since each non-spam message that gets thrown in the 
spam folder undermines the confidence of the user, and decreases the 
likelihood that anti-spam filters will be used universally.  In evaluating the
ease of use of the anti-spam tool, there should be a strong preference for
automated algorithms, which require little or no direct input from individual
users.

We report a surprisingly effective technique that can simultaneously achieve 
the two above-mentioned requirements of accuracy and automation. This 
technique is predicated upon the unique characteristics inherent to social 
networks and the proven wisdom of using such trust networks to make the 
right choices.  Almost all our contractual decisions (e.g., from the schools
we choose for our kids, to the people we trust with our money) depend heavily 
on information provided by our networks of friends.  The reliability of the 
decisions we make, then, depends strongly on the trustworthiness 
of our underlying social networks. Thus, we seem to have evolutionarily 
developed a number of interaction strategies that can generate a 
trustworthy network, and a commonly espoused rule suggests that trust 
is to be built based not only on how  well you know a person, {\em but also} 
on how well that person is known to the other people in your network. This 
interaction dynamic results in ``close-knit" communities; that is, 
communities where, if Alice knows Bob and Charlotte, then it is highly 
likely that Bob and Charlotte also know each other.  In this paper, 
we show that this natural instinct to form close-knit social networks is operative  
 in the cyberspace as well, and can be exploited to provide an  
effective and automated spam filtering algorithm. 

 First, we show that if one constructs personal email networks,
then one can identify distinct close-knit or clustered subnetworks of
email-addresses that communicate with each other via the user, and that
the emails originating from such addresses are trustworthy or non-spam.  
In Sections \ref{sec:construct} - \ref{sec:whitelist}, we discuss 
the construction of personal email networks and the structure of social 
subnetworks.  In Section \ref{sec:blacklist}, we show that there also exist large 
connected subnetworks of email addresses that are {\it not} close-knit, and that
the emails related to these addresses are overwhelmingly spams.  
In Section \ref{sec:efficacy}, we show 
that our algorithm, when used alone, can classify more than half of  all email messages in an individual user's inbox, {\em without any mistakes}. Further analysis shows
that we accurately classified  approximately $44\%$ of {\em all} the non-spam emails, and  
$54\%$ of {\em all spam emails}, while $47\%$ of all email is not
classified, because they are related to subnetworks that are too small in size to allow
reliable determination of statistics. Thus,  the automated 
network-only-based algorithm does not make
any mistakes, but leaves a subset of the messages unclassified; clearly, one needs to use 
more sophisticated content-based filters to classify these. However,   
these complex filters achieve their advertised high accuracies only when trained with a sufficient number of  messages received by individual users. This in turn necessitates manual and intentional user intervention, a potential bottleneck to a widespread deployment of accurate filters.   Since the proposed network-based algorithm automatically generates large samples of both non-spam and spam messages, it can considerably alleviate 
the training problem. Section \ref{sec:use} is devoted to showing how the personal email
networks provide a platform for a comprehensive solution to 
the spam problem, by combining the flexibility of sophisticated,  
content-based filters with the ease of use of an automated filter. 

\section{Personal Email Networks \label{sec:construct}}
\begin{figure}
\begin{center}
\includegraphics[height=1in,width=2in]{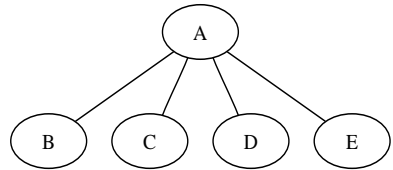}
\caption{The subgraph resulting from an example message from A, to B,C and cc to
D,E.}
\label{fig:exgraph}
\end{center}
\end{figure}

In previous works, email graphs have been constructed based on email
address books\cite{nfb02} or complete email logs of sets of users
\cite{twh03, emb02, ccr03}.  In this work, we build our network based on
the information available to only one user of an email system, specifically,
the headers of all the email messages in that user's inbox.
Each email header contains the email address of one sender, stored in the
``From:" field, and a list of recipient addresses, stored in the ``To:" and
``Cc:" fields.  We construct a personal email network by first creating 
nodes representing all the addresses that appear in all the email headers in 
the user's inbox.  Edges are added between pairs of addresses that appear in the 
same header, i.e. that have communicated via the user. 
For example, suppose Bob sends a message ``To:" Alice and Charlotte, with 
a ``Cc:" to David and Eve, then we can represent this email interaction 
via a star subnetwork, as  illustrated in Fig. \ref{fig:exgraph}.

\begin{center}
\begin{figure*}
\includegraphics[height=4.5in,width=7in]{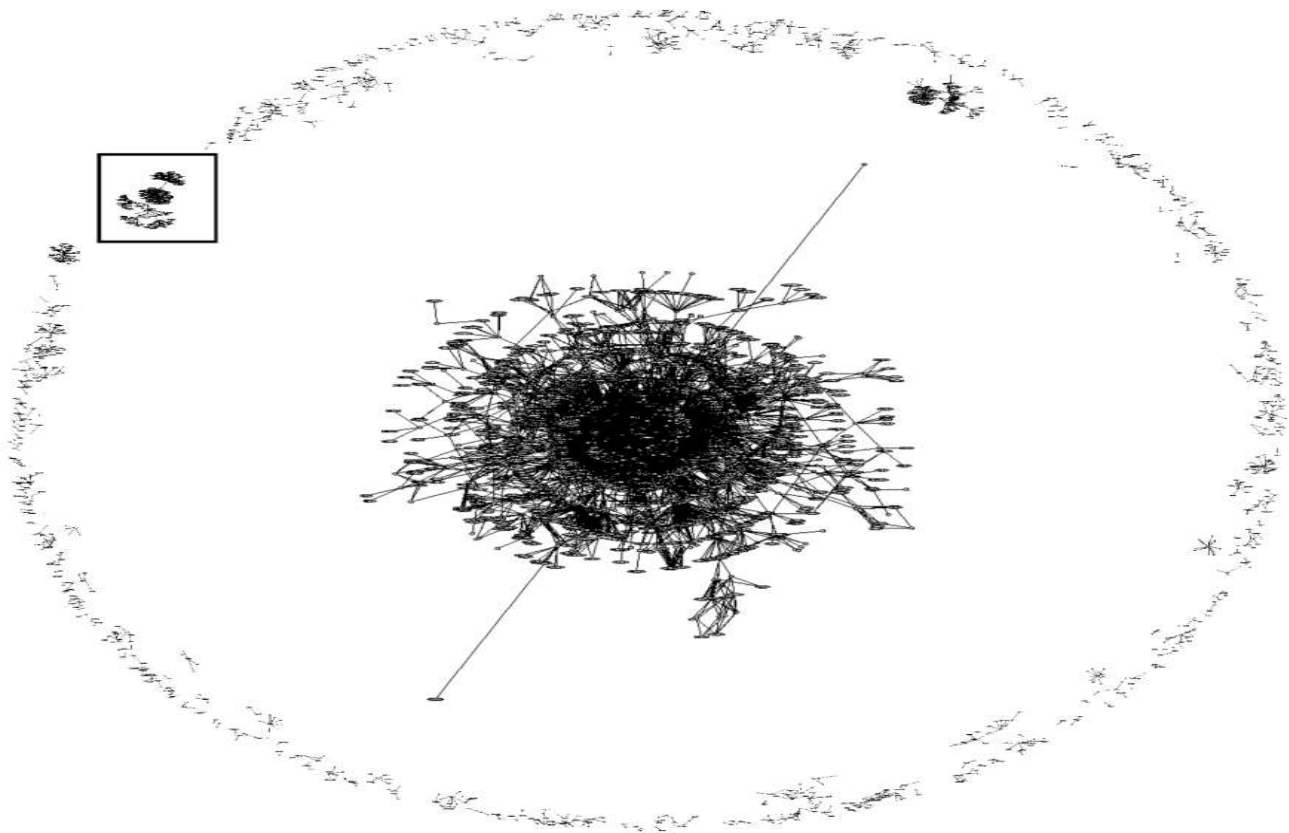}
\caption{One author's complete email network, formed by 5,486 messages.  The largest component
(component 1, at center) is enlarged to show structure.  The second largest, component 2, 
is boxed at upper left.  An enlarged view of component 2 is shown in Fig. \ref{fig:comp2}}
\label{fig:emailnet}
\end{figure*} 
\end{center}

Finally, all nodes representing the user's own email addresses are 
removed, since we are interested only in the connections among email addresses 
who communicate via the user.  Fig.\ref{fig:emailnet} shows a personal email 
network for one of the authors of the paper.  Interestingly, and perhaps 
contrary to intuition, the largest connected component in this particular 
network corresponds to spam-related emails.  This raises the question: how 
does one determine which subnetworks correspond to trusted email addresses 
and which ones correspond to spam-related ones?

\begin{figure}
\begin{center}
\includegraphics[height=3.1in,width=2in,angle=180]{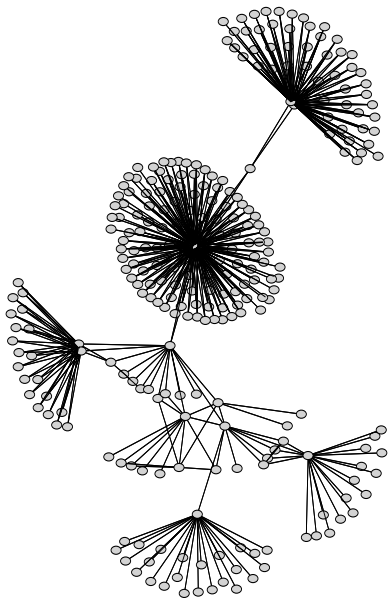}
\caption{The second largest component (called component 2) in one 
author's email graph for a six week period}
\label{fig:comp2}
\end{center}
\end{figure} 

\begin{figure}
\begin{center}
\includegraphics[height=1in,width=2in]{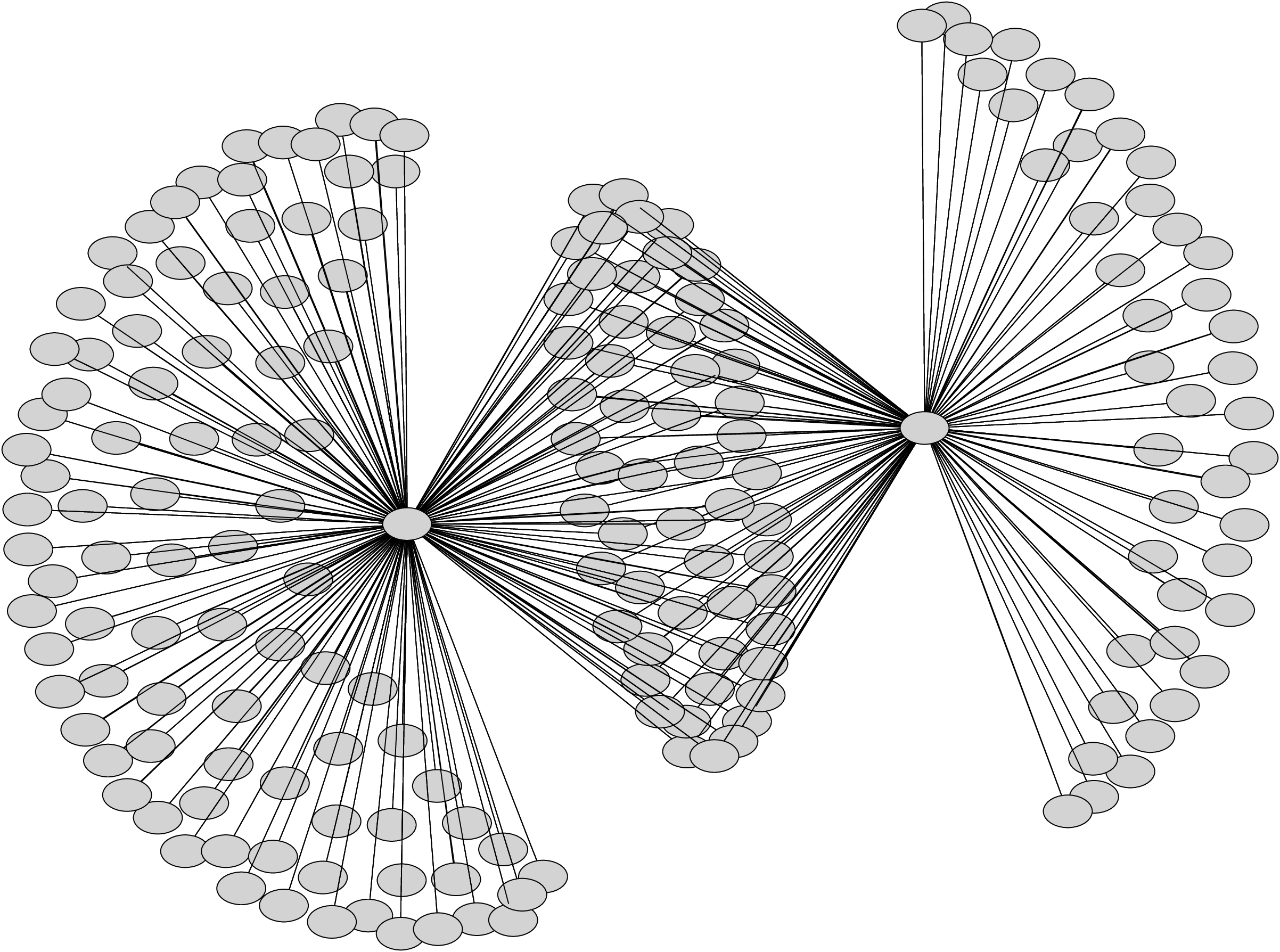}
\caption{A subgraph of a spam component.  See two spammers and their many recipients.
These two spammers share many co-recipients, as seen in the middle of the graph. In 
this subgraph, no node shares a neighbor with any of its neighbors.}
\label{fig:spam-sg}
\end{center}
\end{figure} 

\begin{figure}
\begin{center}
\includegraphics[height=1in,width=1.5in]{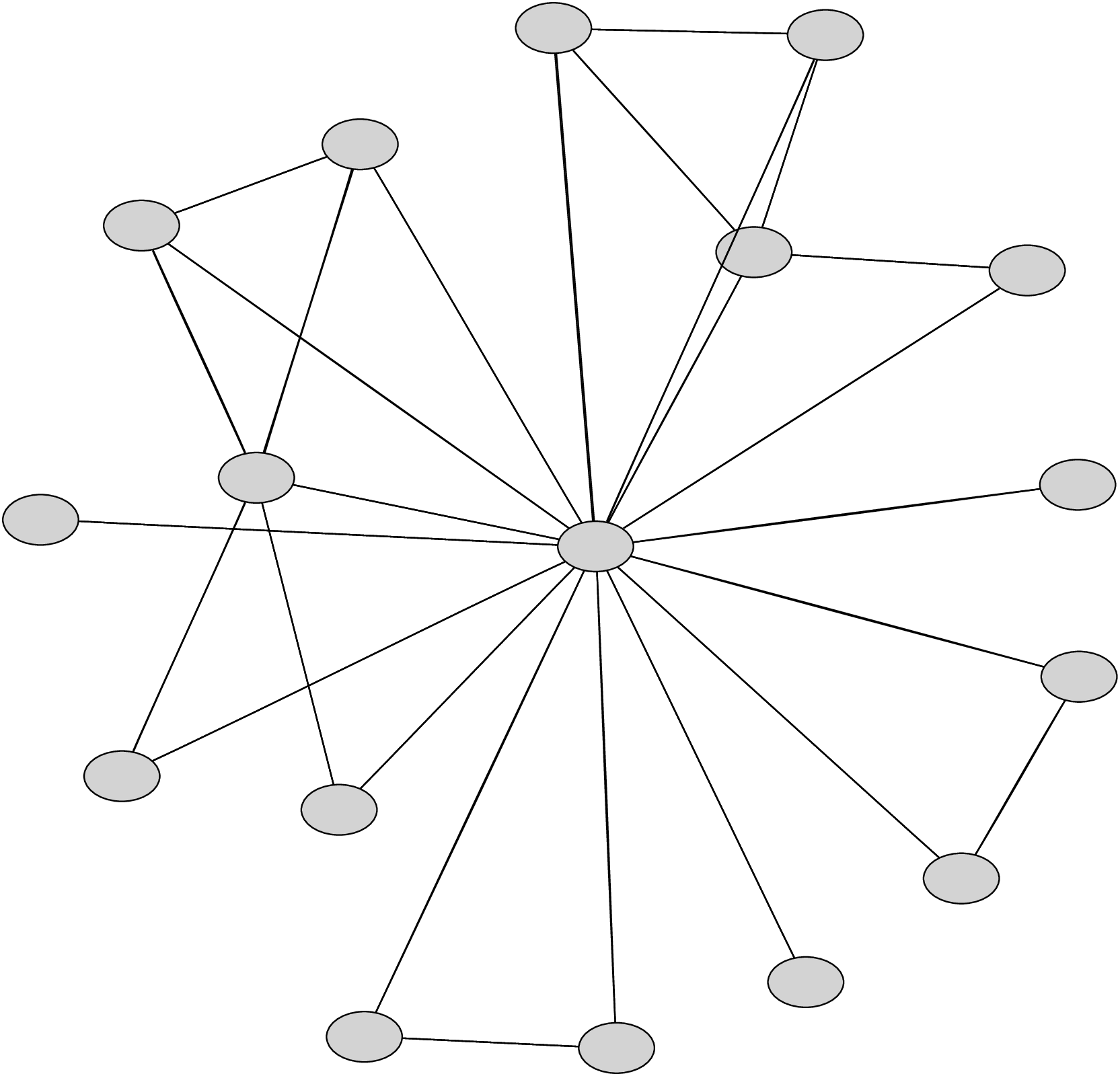}
\caption{A subgraph of a non-spam component.  Note the high incidence of triangle
structures, as compared with the spam subgraph shown in Fig. \ref{fig:spam-sg} }
\label{fig:nonspam-sg}
\end{center}
\end{figure} 

\section{Social Networks and Whitelists \label{sec:whitelist}}
A number of recent studies have identified quantitative measures of the 
closeness of a community, and have shown that, indeed, these measures 
can be used to distinguish empirically observed social networks from 
more anonymous and non-social networks \cite{newmanpark03, newman02, emb02}.
The most distinctive property of social networks is their tendency to
cluster.  For example, if Alice knows Bob and Eve, i.e., 
Bob and Eve are connected through Alice, then,  in a social network, the 
likelihood of Bob knowing Eve is considerably higher than in, for example, 
a random network with very similar degree distribution.  
A qualitative expression for the clustering coefficient of a network is
defined as follows: we count all pairs of nodes that are part of a 
{\em wedge}, i.e., each node in the pair has a direct edge to a third node. 
According to the intuitive notion of clustering 
mentioned above, we expect that in a graph with high clustering coefficient, 
many of these pairs will also be connected via an edge, i.e., many of the 
wedges also form triangles. Hence, the clustering coefficient (sometimes 
called transitivity), $C$, of a graph can be expressed as:
\begin{equation}
\label{eqn:clust_qual}
C = \frac{ 3\times \left(\hbox{number of triangles in the graph}\right)}
{\hbox{number of wedges}} \ 
\end{equation}

This expression gives the reader a feel for the physical meaning
of the clustering coefficient in social networks.  The quantitative definition
of the clustering coefficient that we will use for the rest of this work 
involves counting the fraction of neighbors 
of a node which are also neighbors to each other.  Specifically, if a node 
has degree $k_i$, then it has $k_i$ neighbors.  Each of those $k_i$ neighbors 
potentially are connected to each other.  There are a total of $k_i(k_i-1)/2$ 
possible connections between the neighbors.  By counting the number of 
connections that exist, $E_i$, and dividing by $k_i(k_i-1)/2$,
we get the clustering 
coefficient for the node.  If a node has degree 1, this quantity
is undefined, thus we only count nodes of degree greater than 1.
The clustering coefficient for the entire graph 
is the average of the clustering coefficient for each node (of
degree greater than 1):
\begin{equation}
\label{eqn:clust_quant}
C = \frac{1}{N_2}\sum_{i} \frac{2E_i}{k_i(k_i-1)} 
\end{equation}
where $N_2$ is the total number of nodes in the network with
degree 2 or greater.
This metric has been applied to email graphs previously and has been
found to be more than 10 times larger than one would expect from a random
graph with the same degree distribution\cite{emb02}.

As an example of how clustering coefficient may be used to distinguish between
spam and non-spam emails, consider the connected components 1 and 2, shown in
Fig. \ref{fig:emailnet} and \ref{fig:comp2}, respectively.  Some properties 
of these and other components are listed in Table \ref{tab:emailnet}.  
Component number 1 has a clustering coefficient of 0: exactly zero nodes share
neighbors with any of their neighbors.  On the other hand, component 2, which 
is smaller in size, has a clustering coefficient of 0.67, or, on average, 
67\% of each node's neighbors are connected to each other.  
Figures \ref{fig:spam-sg}
and \ref{fig:nonspam-sg} show subgraphs from components 1 and 2, respectively.  
The relative incidence of the triangle structures that characterize close-knit
communities can be clearly seen in these subgraphs.
Recognizing that social networks have high clustering coefficients, we can be 
confident that the email addresses in component 2 are a part of the user's 
social network in the cyberspace.  Thus, any email with one of the nodes from 
the second component in the header can be classified as a non-spam.  The
email addresses associated with these nodes comprise the user's {\bf whitelist}.

\begin{table}[h]
\begin{tabular}{r|r|r|r|r}
Component & $N$ & $C$ & $k_{max}$ & $\frac{k_{max}+1}{N}$\\
\hline\\
1 & 3560 & 0 & 542 & 0.153 \\
2 & 266 & 0.673 & 112 & 0.425 \\
3 & 174 & 0 & 98 & 0.569 \\
4 & 49 & 0 & 48 & 1 \\
5 & 34 & 0 & 17 & 0.529 \\
6 & 22 & 0 & 21 & 1 \\
\end{tabular}
\caption{Statistics on largest components of complete email network shown
in Fig. \ref{fig:emailnet}.
}
\label{tab:emailnet}
\end{table}

\section{Blacklists and the formation of spam components \label{sec:blacklist}}
Can we also conclude that the first component, which has a very low clustering 
coefficient, is generated by spams?  If so, then any email for which a node 
inside the first component is a sender or co-recipient can be labeled as spam. 
Indeed, a detailed bookkeeping of the inbox shows that the email addresses in 
component 1 are always related to spams, just as those in component 2 are always 
part of the user's social network!  In other words, the email addresses in the 
first component can comprise a {\bf blacklist}.

We have already discussed the reasons why we should expect 
our networks of friends to have a high clustering coefficient, but we have 
yet to understand why a large subnetwork with low clustering coefficient 
should necessarily be spam-induced.  A careful
analysis of component 1 reveals that it has been created by a common form of 
spamming technique called the {\em dictionary attack}, where spammers send messages 
to co-recipients sorted according to email address. For example, if 
one has an email address ``adam@mail.com", one will tend to see co-recipients with
alphabetically similar email addresses, such as ``arthur@mail.com", ``alex@mail.com",
``avid@mail.com", etc.
Due to sorted recipient lists, it is not at all uncommon to have co-recipients 
repeated in different spam email messages, causing the disconnected spam components to 
merge into larger and larger components.
One can view the spam messages as forming a bi-partite graph, with two types of nodes
representing spammers and spam recipients.  Since the spammers don't
spam one another, and the co-recipients of spam messages don't know one another,
this graph will always have a clustering coefficient of 0.

\begin{figure}
\begin{center}
\includegraphics[height=3.2in,width=2in,angle=270]{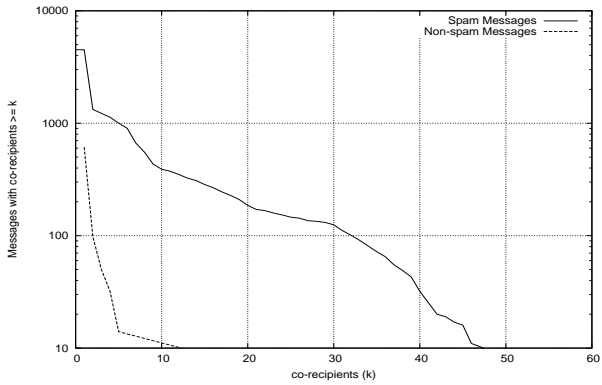}
\caption{CCDF of the number of co-recipients per message for spam and non-spam 
messages.  The x-axis shows the number of co-recipients, and the y-axis is the 
number of messages with at least that number of co-recipients.  The mean number 
of co-recipients for the spam messages is $3.87$; for non-spam, it is $1.71$.}
\label{fig:num_dist}
\end{center}
\end{figure}

An obvious question is: how quickly do we expect the spammer components to merge?
To answer this, we need to examine the probability that two different spammers 
send messages to the same co-recipient.  In figure \ref{fig:num_dist}, we 
see the complementary cumulative distribution function (CCDF) of the number 
of co-recipients per spam message.
In this data, the average number of recipients of a spam message is $3.87$.
As a model, we assume that each spammer uses a ``From:" address only one time,
and sends the spam to $l$ recipients chosen at random.  Based on our data, we 
choose $l\approx 3.87$.  The model assumes that there are $k$ email addresses
near to the user's address in the sorted address space.  

We can solve for the size of the largest connected component in the following way.  
Define $S_i$ to be the size of the largest connected component after $i$ messages.  
The probability that each recipient is already in the largest component is $S_i/k=q$.  
The probability that $m$ of the recipients are in the largest component is
$p_m = {l \choose l-m}(1-q)^{l-m}q^m$.  Putting this together in a rate equation, we 
find:
\begin{eqnarray*}
S_{i+1}&=& \sum_{m=0}^l p_m(S_i + (l - m)) - p_0 l\\
&=& S_i + (1 - p_0)l - q l\\
&=& S_i - \frac{l}{k}S_i + (1 - (1 - q)^l)l\\
\frac{dS_i}{di}&\approx& (1 - (1 - q)^l)l - \frac{l}{k}S_i
\end{eqnarray*}
We can approximate the above in two regimes, namely, unsaturated ($q\ll 1$) and
saturated ($q\approx 1$) to get two different solutions.
In the unsaturated regime,
\begin{eqnarray*}
\frac{dS_i}{di}&\approx& (1 - (1 - q)^l)l - \frac{l}{k}S_i\\
&\approx& l^2 q - \frac{l}{k}S_i\\
&=& \frac{l(l-1)}{k}S_i\\
\Longrightarrow S_i &=& l e^{\frac{l(l-1)}{k}i}
\end{eqnarray*}
In the saturated regime, 
\begin{eqnarray*}
\frac{dS_i}{di}&\approx& (1 - (1 - q)^l)l - \frac{l}{k}S_i\\
&\approx& l - \frac{l}{k}S_i\\
\Longrightarrow S_i &=& k(1 - e^{-\frac{l}{k}i}) + l e^{-\frac{l}{k}i}
\end{eqnarray*}

Clearly, after $O(k / l^2)$ messages, the spam components should
start to join.  This analysis underestimates the joining rate because it always
assumes that a message only joins with the largest component, and never adds
more than one component together, which clearly only increases the rate that the
spam components grow in size.  Additionally, we have ignored the fact that the
nearer the address by alphabetical measure, the more likely they are to be
co-recipients of an email message.  Instead we have approximated that there are
$k$ nearby addresses, and they are all selected at random.

Spammers, of course, will try to defeat any and all anti-spam tools. 
There are a few countermeasures that spammers could take to attempt to
foil the proposed algorithm.  The most obvious countermeasure is to never 
use multiple recipients in the To: or Cc: headers of a spam message.  
In that way, spam components would be isolated nodes in our graph, and
would be disregarded for the purposes of constructing the blacklist.
However, this {\em would not} impact the ability of users to automatically 
generate whitelists, which are extremely valuable in improving the accuracy of
content-based filtering systems. The spammer could also attempt to make the 
algorithm misclassify them as a non-spammer.  For instance, spammers could 
try to learn email addresses for friends of each user by means of email 
viruses, proximity of email addresses on web pages, or other means.  The
spammer could then include the user's friends as co-recipients of a spam
message, thus posing as a member of the user's social network.
However, if spammers can impersonate members of your whitelist they 
can damage the effectiveness of any whitelisting scheme, not just our 
algorithm.  

\section{Greylists: The algorithm and its efficacy \label{sec:efficacy}}

\begin{figure}
\begin{center}
\includegraphics[height=3in,width=1.5in,angle=270]{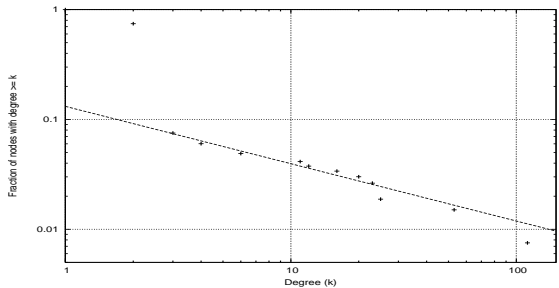}
\caption{Degree CCDF for the largest non-spam component.  The line
corresponds to $p_k\propto k^{-1.52}$.  Note that nodes with degrees 1 
and 2 do not fit the power-law of the tail.  We believe this is due to 
the fact that the personal email network construction can only add an edge 
when the user is a recipient and does not see the edges that would exist 
between two users when all users' email traffic is considered.}
\label{fig:nsdegs}
\end{center}
\end{figure}

We have described an algorithm that divides a single user's personal email network into
disconnected components and attempts to identify each component as spam or non-spam
based only on the value of the clustering coefficient.  We've discussed this metric for
two large subnetworks of one author's email graph, but now we must ask: is it possible to 
classify {\em all} components using this scheme?  For each component, we note the size, 
the maximum degree $k_{max}$, and the clustering coefficient.  We expect that
there is a minimum number of nodes for which the clustering coefficient of
a component can be reliably measured.  Components smaller than this cutoff
size, $S_{min}$, then, must be excluded from our classification scheme.
We also know that, for power-law graphs with exponents greater than or
equal to $-2.0$, we can expect the maximum degree to be on order of
the size of the graph\cite{newmanpark03}.
Previous works \cite{emb02, ccr03}, as well as this one (see Fig.
\ref{fig:nsdegs}), find degree distributions with exponents greater
than $-2.0$.
We use this fact to
introduce another cut-off parameter.  All components with clustering coefficient 
equal to $0$ and $(k_{max}+1)/size$ greater than $K_{frac}$, are also disregarded,
in order to limit the impact a single node can have on the statistics.
Remaining components with clustering coefficient less than some critical value, 
$C_{min}$, are assumed to be spam components, and all nodes in these components 
are written to the blacklist.  If a component has clustering coefficient greater 
than $C_{max}$, then we write all nodes to the whitelist.
The rare case in which a component which was not disregarded has a clustering 
coefficient {\em between} $C_{min}$ and $C_{max}$ is addressed later in this section.

The criteria for choosing the cutoff
parameters $S_{min}$, $K_{frac}$, $C_{min}$, and $C_{max}$ are as follows:
\begin{itemize}
\item {\em Choice of $S_{min}$ and $K_{frac}$:}
Single messages can make isolated components which it is not clear how to classify 
a priori, since every message with $k$ co-recipients can create a component with 
clustering coefficient equal to $0.0$ and size $k$.  Setting $K_{frac}$ less than $1.0$ 
($0.6-0.8$ works well in practice) ensures that a component from a single 
message will not be considered.  A more direct route (but one that can't
be realized using purely graph theoretic methods) is to only consider components
formed by $N$ or more messages.
The size cutoff
should come from the number of recipients a message is expected to have.
Setting this far above the mean also insures that several messages are
required for each component.  For the data we examined, a cutoff size
of $10-20$ seemed to work well.

\item {\em Choice of $C_{min}$ and $C_{max}$:} As we saw in section
\ref{sec:blacklist}, we expect spam components to have $C=0$.  On the
other hand, in our data, as in previous studies\cite{emb02,newmanpark03},
the clustering coefficient of the social graphs is found to be an order of
magnitude larger than one would expect for a random graph with the
same degree distribution.  
In this work, $C_{min}=0.01$, and $C_{max}=0.1$ produce excellent results.
\end{itemize}

The methods for choosing the various parameters could be improved with further
research on more personal email networks to better understand the
statistical properties of these networks over a larger set of users.

We have already discussed how email addresses associated with components 
larger than $S_{min}$ and with clustering coefficients less than $C_{min}$ 
or greater than $C_{max}$ are sorted into whitelists and blacklists.   We
now resolve how to deal with large components that have an intermediate value
of the clustering coefficient.  

In general, we assume that spammers don't know who the user's friends are, 
so it is \emph{very} unlikely for a spammer to email the user and a friend 
of the user in the same message. 
This presumably\footnote{assuming spammers do not do surveillance to learn 
the structure of social networks} happens purely by chance, and is also 
presumably very rare, since we can imagine that each co-recipient on a spam
message is a friend with some very small probability.  
Hence, spam components usually stay disconnected 
from friend components.   There is always the possibility, however, that a 
spam message will have a co-recipient in common with a non-spam message.  
The cross-component 
connections are most likely to happen in spam components with a large number 
of co-recipients; this means that chance connections will result in a non-spam 
component joined with a large spam component via a small number of edges (the 
chance co-recipient connections). 
From a graph theoretic perspective, this situation will correspond to two 
connected ``communities" which have very few edges between them, and which 
also have very different clustering coefficients.
We may identify these cases when we find large components with intermediate
clustering coefficients.  For instance, if the component size is large, but
the clustering coefficient is less than $C_{max}$, we may assume that it is
a joined spam component.  

A metric called {\em edge betweenness} has been suggested
as a tool to identify edges that go between communities
\cite{twh03, ng03, newman0309}
The betweenness of an edge within a network is a measure of how many
of the shortest paths between all the pairs of nodes in the network
include that edge.  Since all paths linking
the many nodes in the spam community to nodes in the non-spam community in 
a joined component will include one of the very few edges that correspond 
to the chance connections between the two communities, these edges 
will clearly have a much higher betweenness than the edges 
that connect members within the same community.
Therefore, in our approach, we split joined components into two communities 
by removing the edges with the highest betweenness until there are two 
distinct components, recalculating the betweenness after each edge removal.  
This step is the same as the algorithm given by Newman
and Girvan\cite{ng03}; however, we do not execute this step on components with high
clustering coefficient (in fact, Newman and Girvan's community-finding algorithm 
tends to cut these email graphs into many communities, whereas we are interested 
only in finding the split between spammers and non-spammers).

When considering a data set from one author of all the saved email messages
over a period of 3 years,
we found only one example where an edge arose between a spam
and non-spam community.  This edge was indeed removed by the above separation
technique.
Therefore, we expect that this separation technique 
will rarely be needed, but we also expect that it is a robust method for separating
joined components, as long as spammers do not have access to the user's whitelist.

\begin{table}[h]
\begin{tabular}{r|r|r|r|r}
Data & blacklist & whitelist & greylist & total\\
\hline\\
spam 1 & 1,664 & 0 & 2,841 & 4,505\\
non-spam 1 & 0 & 331 & 282 & 613 \\
spam 2 & 2,981 & 0 & 1,113 & 4,114 \\
non-spam 2 & 0 & 68 & 229 & 297
\end{tabular}
\caption{Results of the algorithm on two datasets.  Data set
1 is from a 6 week period, set 2 is from a 5 week period.
The data sets are from two different users.}
\label{tab:results}
\end{table}

In Table \ref{tab:results}, we see the results of the algorithm
on two data sets covering 5 and 6 weeks for two different users.
Averaging over both sets, $44\%$ of the nonspam is on the whitelist,
$54\%$ of the spam is on the blacklist, and $47\%$ of the 
messages are on the greylist.  In summary, about half the time, the message is
correctly identified, and the other half, the algorithm cannot classify the
message.  Over the test data, not one message is misclassified.

It is interesting to note that our algorithm may be improved by 
adding more graph theoretic parameters to our classification scheme.
Figure \ref{fig:nsdegs} shows the CCDF of the largest non-spam component.
The degree distribution for the tail of this CCDF is $p_k\propto k^{-1.52}$.
The nodes with degrees 1 and 2 do not fit the power-law of the tail.
It is interesting to note that this approach of taking a user's view to make
a subgraph of the entire email network gives a degree distribution
consistent with previous works\cite{emb02, ccr03}, which were based on the 
email of multiple users.  These previous studies found degree
distributions following power laws with exponents from $-1.3$ to $-1.8$.
In our datasets, the spam components had power-laws with exponents
between $-1.8$ and $-2.0$.  Thus, since spam components seem to have 
unusually high power law exponents, it may be possible to use the degree distribution
of a component to improve the distinguishing power, or reduce the likelihood
of error.

\section{How to use the algorithm? \label{sec:use}} 
As already discussed, our email-network-based spam filtering algorithm automatically 
constructs whitelists with $44\%$ success rate, constructs blacklists with 
$54\%$ success rate with no false classifications, and leaves $47\%$ of the
email unclassified.  These hit rates are achieved without any user intervention.
Since the only information necessary is available in the user's email headers, 
the algorithm can be easily implemented at various points in the filtering process.
Since whitelists and blacklists are the 
foundation of many anti-spam approaches\footnote{Existing whitelist systems 
use a list of email addresses or mail servers to accept mail from, and all 
senders not on the whitelist are assumed to be spam.  Similarly, 
existing blacklist systems have a list of email addresses or mail servers 
to block mail from, and all senders not on the list are assumed to be non-spam. 
These lists are currently created in an ad-hoc manner, based on global prior 
knowledge and/or user feedback.}, our technique, based purely on graph 
theoretic methods, can be easily integrated with existing methods.  
For instance, our algorithm may be used by mail administrators of large email 
servers, such as corporate email servers, or Internet service providers, since
they can generate a personal email network for each of their users as the
mail is being delivered to their mail servers.
The ability to generate whitelists and blacklists for all their users will 
greatly improve the ability of central mail servers to reduce the number of 
spam messages which reach end users. 

Another major advantage of the algorithm is that it is virtually immune to 
false negatives (e.g., non-spam being identified as spam), which suggests 
that our method can {\em also} significantly increase the ease of use of 
content-based anti-spam tools, which classify emails as spam or non-spam 
based on the content of the email, rather than the sender's address.
The best content-based filters achieve approximately 99.9\% accuracy, but 
require users to provide a training set of spam and non-spam messages.  
These algorithms examine the content of the sets of 
messages labeled as spam or non-spam to identify common characteristics 
of the two types of mail, often using Bayesian inference\cite{crm114}.
The performance 
of content-based algorithms is usually greatly improved if the training set
comes from emails sent to the user who will run the algorithm, which is to
say, these algorithms work better when tuned to specific email inboxes.
Our method can \emph{automatically} generate an accurate training set, tailored 
to the individual user's inbox, so that the end user can be less involved 
in the training process.  This could dramatically impact the ease of widely 
deploying content-based anti-spam tools, which have been heretofore 
inconvenient for end users to train effectively.  Future research could 
produce more sophisticated schemes to better combine our method with 
content-based learning methods.

\section{Concluding Remarks}
We have proposed an algorithm based on the properties of social networks to
classify senders as spammers or non-spammers.  The algorithm has not
misclassified {\em any} senders in the email data sets we have tried.  
Using this algorithm,  it would be possible to generate training sets for 
learning algorithms which may be content-based.  This is extremely 
important in achieving accurate, automated spam filtering.

There are many areas for further improvement of the algorithm.  Most
obviously, more parameters of components should be considered.  In this paper,
we classify components based only on size and clustering coefficient, but there 
are many more metrics which may provide for statistical distinguishability 
between spam and non-spam.
An algorithm which incorporates more parameters of the
graph may be able to reduce the probability of misclassification even further.
Unfortunately, significantly decreasing the size of the greylist
does not appear very promising since the greylist components are very small in size,
and thus live in a small parameter space, and hence, strong distinguishability
between greylisted components appears to be beyond the power of purely graph-based
techniques.
As long as spammers 
continue to adapt their strategies for defeating anti-spam tools, improving 
these tools is a subject that will warrant further study.

In the long run, there is clearly a solution to the spam problem: use 
cryptographic whitelists, where a user only accepts messages that are
cryptographically signed by authenticated keys. The problem, of course, 
is that the lack of infrastructure that is needed to make this scheme 
accessible to the majority of end users makes the immediate potential for 
widespread use of such a scheme highly questionable.  Until this infrastructure
is in place, end users must rely on less perfect solutions, and will benefit 
from any effort that makes these solutions more user-friendly, and easier for
mail servers and ISPs to broadly distribute.

Moreover, even if the spam problem is solved, the email network tool introduced 
here will become increasingly useful. In particular, the personal
email network has the potential to enable users to capture the community structure
in cyberspace.  It is possible that better email message management can be achieved
if email clients are aware of the various social groups of which the user is a
member, and our scheme for generating personal email networks may give such
information without any changes to Internet email protocols.

The authors would like to thank Jesse Bridgewater for use of his
email, and Joseph Kong and Kamen Lozev for their comments.
The authors particularly thank Julie Walters for her many
helpful comments and suggestions.

\bibliographystyle{plain}
\bibliography{networks}
\end{document}